\documentclass[prl,aps,epsfig,superscriptaddress]{revtex4}
\usepackage{bm}
\usepackage{amsfonts}
\usepackage[dvips]{graphicx}
\usepackage{mathrsfs}
\usepackage[intlimits]{amsmath}
\usepackage[colorlinks, citecolor=red]{hyperref}

\begin{document}

\title{Quantum Communication between Multiplexed Atomic Quantum Memories}
\author{C. Li$^{}$}
\affiliation{Center for Quantum Information, IIIS, Tsinghua University, Beijing 100084, PR China}

\author{ N. Jiang$^{\footnotemark[1]}$}
\affiliation{Center for Quantum Information, IIIS, Tsinghua University, Beijing 100084, PR China}

\author{Y.-K. Wu$^{}$}
\affiliation{Center for Quantum Information, IIIS, Tsinghua University, Beijing 100084, PR China}

\author{W. Chang$^{}$}
\affiliation{Center for Quantum Information, IIIS, Tsinghua University, Beijing 100084, PR China}

\author{Y.-F. Pu$^{\footnotemark[2]}$}
\affiliation{Center for Quantum Information, IIIS, Tsinghua University, Beijing 100084, PR China}

\author{S. Zhang$^{}$}
\affiliation{Center for Quantum Information, IIIS, Tsinghua University, Beijing 100084, PR China}

\author{L.-M. Duan$^{\footnotemark[3]}$}
\affiliation{Center for Quantum Information, IIIS, Tsinghua University, Beijing 100084, PR China}

\renewcommand{\thefootnote}{\fnsymbol{footnote}}
\footnotetext[1]{Present address: Department of Physics, Beijing Normal University, Beijing 100875, China}
\footnotetext[2]{Present address: Institute for Experimental Physics, University of Innsbruck, A-6020 Innsbruck, Austria.}

\begin{abstract}
The use of multiplexed atomic quantum memories (MAQM) can significantly enhance the efficiency to establish entanglement in a quantum network. In the previous experiments, individual elements of a quantum network, such as the generation, storage and transmission of quantum entanglement have been demonstrated separately. Here we report an experiment to show the compatibility of these basic operations. Specifically, we generate photon-atom entanglement in a $6\times 5$ MAQM, convert the spin wave to time-bin photonic excitation after a controllable storage time, and then store and retrieve the photon in a second MAQM for another controllable storage time. The preservation of quantum information in this process is verified by measuring the state fidelity. We also show that our scheme supports quantum systems with higher dimension than a qubit.
\end{abstract}

\maketitle

Quantum network is one of the central targets of quantum information science \cite{kimble2008quantum}, with wide applications in quantum communication \cite{PhysRevLett.67.661,PhysRevLett.70.1895} and distributed quantum computing \cite{PhysRevA.59.4249}. To generate and distribute quantum entanglement among distant nodes of a quantum network, the idea of a quantum repeater \cite{briegel1998quantum} is proposed and it is shown that high-fidelity quantum memories are necessary for its efficient implementation \cite{duan2001long,sangouard2011quantum}.

The atomic ensemble has been one of the most popular candidates for realizing quantum repeaters since the proposal of the DLCZ protocol \cite{duan2001long}: simply using linear optics and photon counting technology, the quantum information can be stored as atomic excitations in the ensemble with long lifetime, and can be efficiently retrieved as flying photons through the collective effect of the atoms \cite{hammerer2010quantum}. Tremendous progress has been made in this field: the basic elements of a quantum network, the generation, transmission, storage and retrieval of quantum information have been achieved in the atomic ensemble \cite{van_der_Wal196,Matsukevich663,chaneliere2005storage} and the entanglement distribution between remote ensembles have been demonstrated \cite{chou2005measurement,Chou1316}.

To further improve the efficiency of entanglement distribution, many variants of the DLCZ protocol have been proposed \cite{sangouard2011quantum}. Among them is the use of multiplexed quantum memories \cite{collins2007multiplexed}: with the ability to entangle arbitrary pairs of memory cells in distant nodes, multiplexing can outperform simple parallelization and significantly reduce the communication time, especially when the coherence time of the memory is limited. Pioneering experiments have demonstrated multiplexing in the atomic ensemble and its solid-state variant using the
spatial \cite{lan2009multiplexed, pu2017experimental}, angular \cite{Chrapkiewicz2017highcapacity} or temporal modes
\cite{saglamyurek2011broadband,Usmani2010mapping,Tang2015storage,
Laplane2016Multiplexed}. For the spatial multiplexing, fundamental elements of the quantum network have been realized separately, such as the heralded generation of atomic excitation \cite{pu2017experimental} and entanglement \cite{pu2018experimental}, their conversion to photonic qubits, and the storage and retrieval of the photonic qubits in arbitrary memory cells \cite{Jiang2019Experimental}.

With these individual elements at hand, it is still important to show their compatibility in a single setup \cite{chaneliere2005storage}. Therefore, in this experiment, we first combine these operations together and demonstrate the generation of photon-ensemble entanglement in a multiplexed quantum memory, the quantum state transfer from atomic qubits to time-bin photonic qubits, and further the storage and retrieval of the photonic qubits in a second multiplexed memory. Then we show that our experimental setup has a native support for high-dimensional quantum systems (qudits), which can enhance the efficiency of various tasks of the quantum network like quantum communication \cite{PhysRevA.61.062308,PhysRevA.64.012306,PhysRevLett.88.127902,PhysRevLett.96.090501,Islame1701491} and quantum computing \cite{1707.08834,1905.10481}.

\begin{figure}[ptb]
  \centering
  \includegraphics[width=18cm]{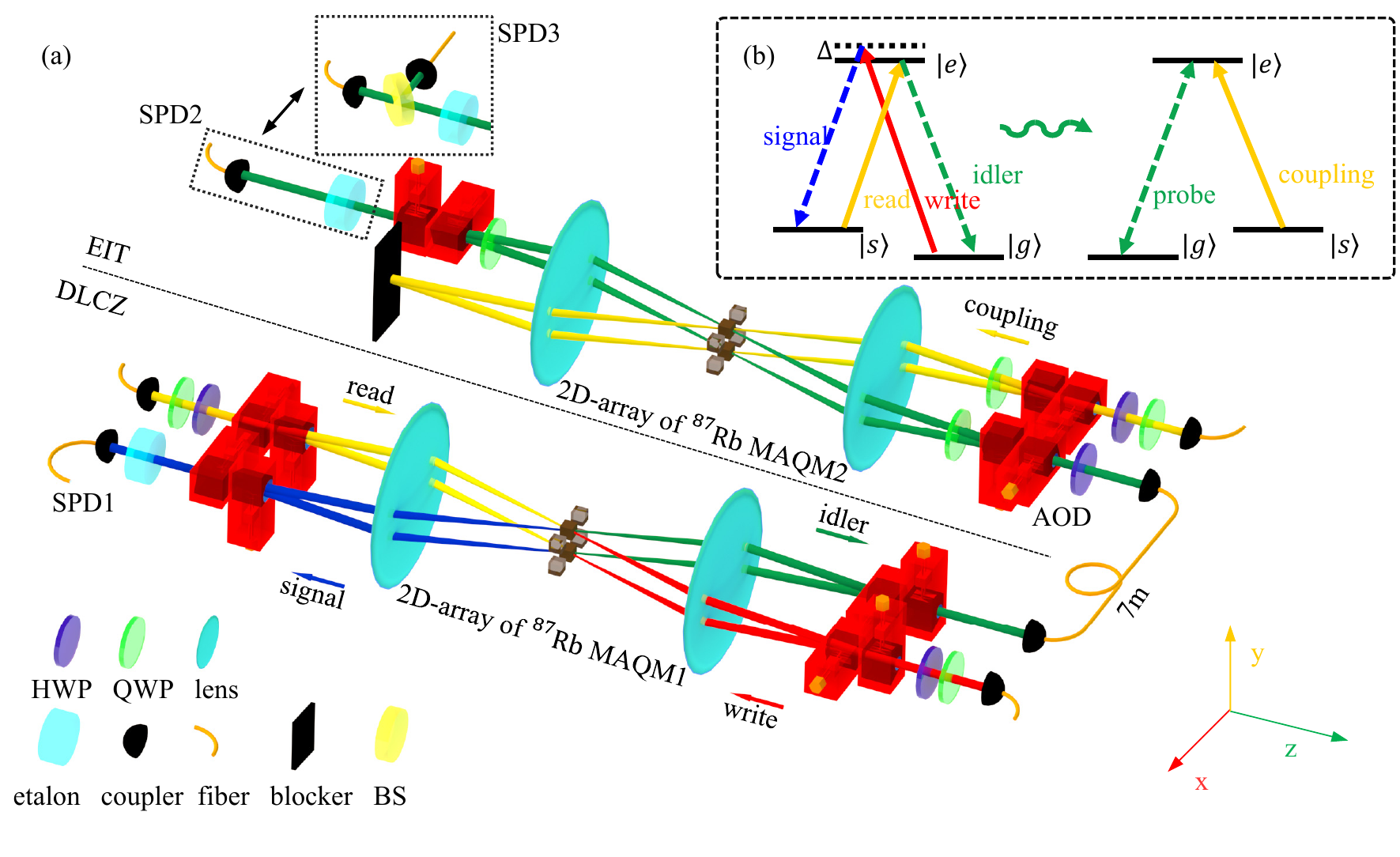}\\
  \caption{\textbf{Generation, transmission and storage of entanglement in multiplexed atomic quantum memory (MAQM).} \textbf{(a)} Simplified setup. HWP represents half-wave plate, QWP for quarter-wave plate, BS for beam splitter and SPD for single photon detector. The photon-atom entanglement is generated by the DLCZ protocol in $^{87}$Rb MAQM1, and stored by EIT in MAQM2. The two MAQMs are connected by a $7\,$m fiber. The write/read beams are counter-propagating, so are the signal/idler photon modes; while the two pairs have an angle of $1.5^{\circ}$. Similarly, the probe mode and the coupling beam of MAQM2 intersects at the same angle of $1.5^{\circ}$. The Gaussian waists of the signal/idler, write/read, probe and coupling modes are $90\,\mu$m, $195\,\mu$m, $140\,\mu$m and $200\,\mu$m, respectively; and the polarizations of the write, the read and the coupling beams are vertical, horizontal and left-handed circular, respectively. The distance between two adjacent micro-ensembles is $300\,\mu$m for both MAQM1 and MAQM2. A beam splitter and an SPD3 can be inserted to measure the non-classical property of the signal photon and the spin-wave excitation (see Supplementary Materials). \textbf{(b)} Energy diagrams for the write, the read, and the EIT storage processes. $|g\rangle\equiv|5S_{1/2},\,F=1\rangle$, $|s\rangle\equiv|5S_{1/2},\,F=2\rangle$, $|e\rangle\equiv|5P_{1/2},\,F=2\rangle$. The write beam is blue detuned by $\Delta=18\,$MHz to the resonant transition at the central memory cell.}
\end{figure}

\section{Results}
\subsection{Generation of photon-atom qubit entanglement and quantum state transfer}
The experimental setup is shown in FIG.~1, which consists of two cold $^{87}$Rb ensembles in two magneto-optical traps (MOT). All the atoms are initially prepared in the ground state $|g\rangle \equiv |5S_{1/2},F=1\rangle$. A weak write beam is applied to generate the quantum correlation between a signal photon and a spin-wave excitation in one atomic ensemble (MAQM1) through the DLCZ scheme. The signal photon is collected by a single photon detector (SPD1); while after a controllable storage time, the spin wave is retrieved by a strong read beam to an idler photon,
which is directed to the second ensemble (MAQM2) by a $7\,$m fiber, and is further stored there as a collective spin wave via electromagnetically induced transparency (EIT).
After a second controllable storage time, this spin wave is retrieved and finally collected by SPD2.

Crossed acousto-optic deflectors (AODs) and lens under $4f$ configuration are used for 2D multiplexing and de-multiplexing \cite{pu2017experimental}. All the beams and single photon modes can be directed to or collected from a particular site of the ensembles by programming the RF signals in the crossed AODs.
In this way, the two atomic ensembles are divided into two $6\times5$ arrays of micro-ensembles, which form our two multiplexed access quantum memories (MAQM). As we show in Supplementary Materials, each memory cell can be addressed individually with low crosstalk errors.
In the following experiment, three pairs and a $2\times2$ sub-array of memory cells in each MAQM are chosen to demonstrate the entanglement generation, transmission and storage (see FIG.~2(a), 2(b)).
\begin{figure}[ptb]
  \centering
  \includegraphics[width=18cm]{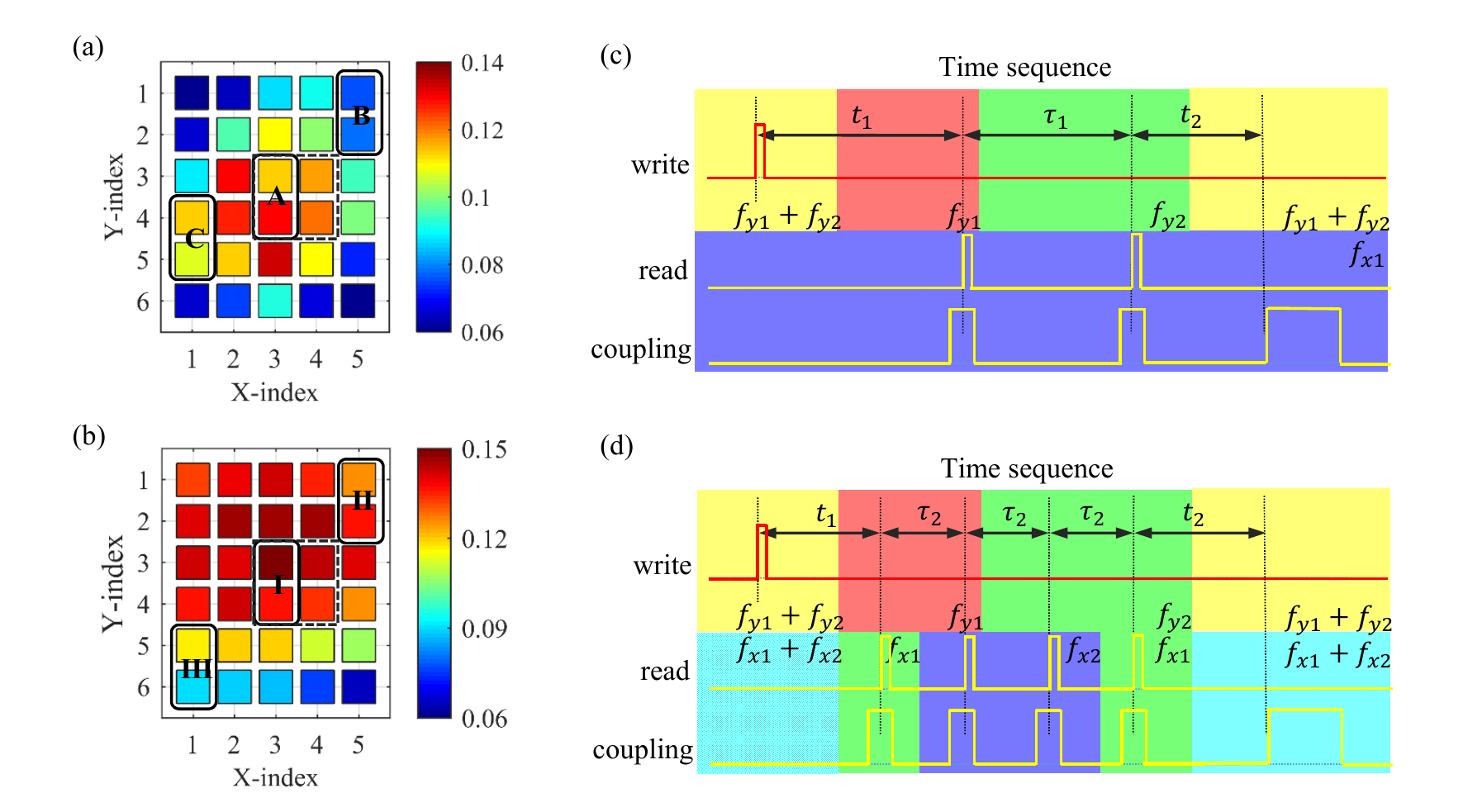}\\
  \caption{\textbf{Retrieval efficiency of MAQMs and control pulse sequences.} \textbf{(a)} The retrieval efficiency of MAQM1. In the 2D array of $5\times6$ memory cells, three pairs of cells (A, B and C enclosed by solid lines) and a $2\times2$ sub-array (enclosed by dashed lines) are chosen to demonstrate qubit/qudit entanglement between the signal photon and the atomic ensemble. The signal photon detection probability is about $1\%$, that is, about 100 write-clean cycles (see Methods) are repeated before one signal photon is detected. The storage time is $15.6\,\mu$s. \textbf{(b)} Retrieval efficiency of MAQM2. In the 2D array of $5\times6$ memory cells three pairs of cells (\uppercase\expandafter{\romannumeral1}, \uppercase\expandafter{\romannumeral2} and \uppercase\expandafter{\romannumeral3} enclosed by solid lines) and a $2\times2$ sub-array (enclosed by dashed lines) are used for EIT storage of the qubit/qudit. To measure the storage efficienty, we apply a weak coherent pulse with a mean photon number of 0.5. The EIT storage time is $7.8\,\mu$s. \textbf{(c)} Time sequence of control pulses for the qubit entanglement transfer. The duration of the write, the read and the coupling pulses are $100\,$ns, $500\,$ns, and $700\,$ns, respectively, and that of the final coupling pulse to readout the spin wave from MAQM2 is $1000\,$ns. The storage times are $t_1=15.6\,\mu$s for the spin wave in MAQM1, and $t_2=7.8\,\mu$s in MAQM2. The interval between the two time bins of the flying qubit is $\tau_1=7.8\,\mu$s. The colored background represents the RF-signal components in all the crossed AODs: the upper row is for the $y$ direction, and lower row for the $x$ direction. \textbf{(d)} Time sequence of control pulses for the qudit entanglement transfer. The storage times are $t_1=11.7\,\mu$s, $t_2=7.8\,\mu$s, and the interval between time bins of the flying qudit is $\tau_2=3.9\,\mu$s. }

\end{figure}

The pulse sequence for the qubit state generation, transmission and storage is shown in FIG.~2(c).
When the write beam is equally split into two micro-ensembles located at $(x_1,\,y_1)$ and $(x_1,\,y_2)$, conditioned on a signal photon being detected, the state of the atomic ensemble and the signal photon prior to the detection can be described by
\begin{equation}
|\Psi\rangle = \frac{1}{\sqrt{2}} \left(|x_1,\,y_1\rangle_{s}|x_1,\,y_1\rangle_{a_1} + e^{i\phi}|x_1,\,y_2\rangle_{s}|x_1,\,y_2\rangle_{a_1}\right),
\end{equation}
where subscripts $s$ and $a_1$ denote the signal photon and the spin-wave excitation in MAQM1, and the relative phase $\phi$, which we set as zero in this experiment, can be precisely controlled by the phase of the RF signal in the AODs.

After a controllable storage time $t_1=15.6\,\mu$s in MAQM1, we retrieve the spin-wave excitation into a photonic qubit. Note that the spin wave in the two memory cells should not be retrieved out at the same time, otherwise the two idler modes will interfere at the demultiplexing AODs that combine them together. Instead, we convert the spin-wave qubit into a time-bin qubit
and transfer it to MAQM2 for storage.
First, we switch all the AODs to address the $(x_1,\,y_1)$ cell in MAQM1 and the $(x'_1,\,y'_1)$ cell in MAQM2. The idler photon is retrieved from the $(x_1,\,y_1)$ cell of MAQM1 by a strong read beam, and is further stored into the $(x'_1,\,y'_1)$ cell of MAQM2 by adiabatically shutting off the coupling beam. Then the state is described by
\begin{equation}
|\Psi'\rangle = \frac{1}{\sqrt{2}} \left[e^{i(\alpha_1-\beta_1)}|x_1,\,y_1\rangle_{s}|x'_1,\,y'_1\rangle_{a_2} + |x_1,\,y_2\rangle_{s}|x_1,\,y_2\rangle_{a_1}\right],
\end{equation}
where $\alpha_1$ and $\beta_1$ are the phases introduced by the read and the coupling beams respectively. The additional phase due to the transmission of the photon is fixed for the given memory cells and thus can be absorbed into the definition of the basis states. Followed by the same operations on the $(x_1,\,y_2)$ cell in MAQM1 and the $(x'_1,\,y'_2)$ cell in MAQM2 after an interval of $\tau_1=7.8\,\mu$s between the time bins, the state becomes
\begin{equation}
|\Psi''\rangle = \frac{1}{\sqrt{2}} \left[e^{i(\alpha_1-\beta_1)}|x_1,\,y_1\rangle_{s} |x'_1,\,y'_1\rangle_{a_2} + e^{i(\alpha_2-\beta_2)}|x_1,\,y_2\rangle_{s} |x'_1,\,y'_2\rangle_{a_2}\right].
\end{equation}
In the experiment, the read and the coupling beams are produced from the same laser, with the coupling beam being guided to MAQM2 by an additional $7\,$m fiber (not shown in FIG.~1). The relative phase fluctuation between the two paths is small during the experimental cycle even without any active phase-locking technique, and the wavefront of the two beams are within the coherence time as the spin-wave excitation is transferred from MAQM1 to MAQM2. Consequently, the phase factors in the above equation can be cancelled and we get
\begin{equation}
|\Psi''\rangle = \frac{1}{\sqrt{2}} \left(|x_1,\,y_1\rangle_{s} |x'_1,\,y'_1\rangle_{a_2} + |x_1,\,y_2\rangle_{s} |x'_1,\,y'_2\rangle_{a_2}\right).
\end{equation}
Finally, by turning on the coupling beam again after a controllable storage time of $t_2=7.8\,\mu$s, we retrieve the spin-wave excitation in MAQM2 for single-photon measurements.

The qubit entanglement is verified by quantum state tomography \cite{James2001On}. In FIG.~3(a) we show the reconstructed density matrix of the signal photon and the atomic qubit in pair A of MAQM1 by blocking the coupling, cooling and repumping beams of MAQM2; and the corresponding reconstructed density matrix after EIT storage in pair \uppercase\expandafter{\romannumeral1} of MAQM2 is shown in FIG.~3(b). Entanglement fidelity $F=\langle \Psi_0|\rho |\Psi_0\rangle$ can be used to quantify the entanglement when the atomic
qubit is stored in MAQM1 or MAQM2
, where $|\Psi_0\rangle$ is the maximally entangled two-qubit state and $\rho$ the reconstructed experimental density matrix. In Table 1 we report these values for quantum state transfer between several typical pairs of memory cells. All the measured entanglement fidelities are above 0.86 in MAQM1, and only decay slightly after the transmission to MAQM2. The decay of fidelity is mainly caused by the small difference in the EIT storage-retrieval efficiency of the two micro-ensembles. We also compute the transmission fidelity, the similarity between the entangled states when the atomic qubit is stored in MAQM1 and MAQM2, for these cases, which are all above 0.85.

\begin{figure}[ptb]
  \centering
  \includegraphics[width=18cm]{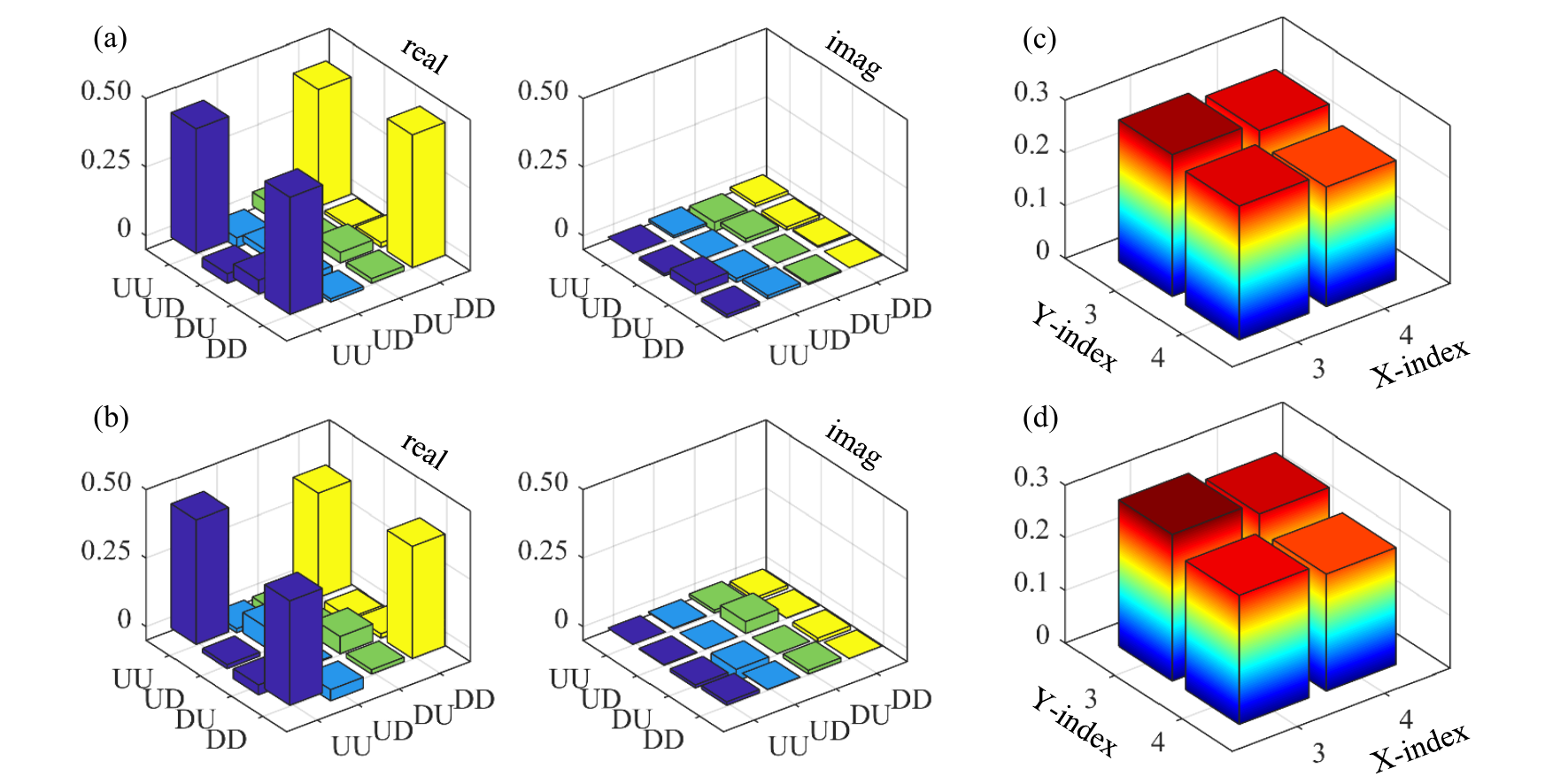}\\
  \caption{\textbf{Entanglement verification.} The reconstructed density matrices for the signal photon and the atomic qubit in \textbf{(a)} pair A of MAQM1 and \textbf{(b)} pair \uppercase\expandafter{\romannumeral1} of MAQM2. The $U$ and $D$ represent the two memory cells  $(x_1,\,y_1)$ and $(x_1,\,y_2)$ respectively. The measured spin-wave excitation population in each of the $2\times2$ sub-array in \textbf{(c)} MAQM1 and \textbf{(d)} MAQM2. Note that they do not characterize the coherence between different memory cells, which we measure by the W-state fidelity in Table~1.}
\end{figure}
\begin{table}[ptb]
  \centering
  \includegraphics[width=18cm]{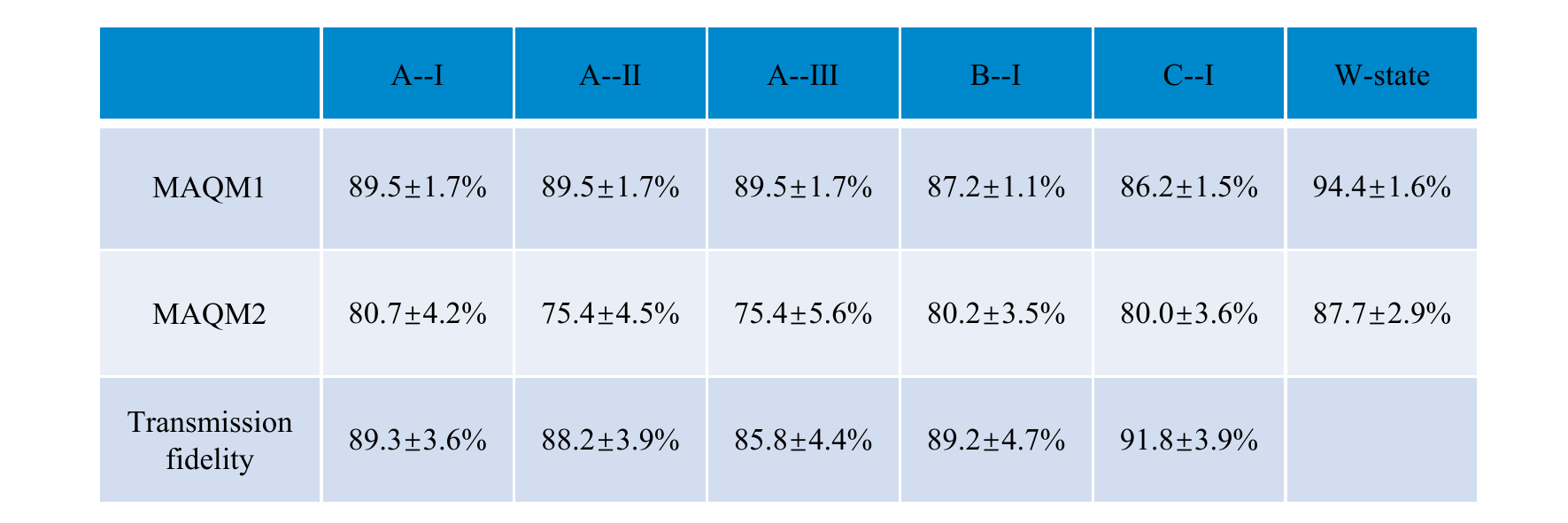}\\
  \caption{\textbf{Fidelity of quantum states.} The entanglement fidelity of the qubit state is calculated by comparing the reconstructed density matrix with the ideal maximally entangled state. In the third row we directly compute the fidelity between the two reconstructed density matrices when the atomic qubit is in MAQM1 and in MAQM2; the high fidelity shows a faithful transmission of the quantum state. The qudit state transmission is characterized by the W-state fidelity \cite{pu2018experimental}; here the transmission fidelity is not computed because we do not reconstruct the complete density matrices. The error bars are calculated by the Monte Carlo simulation with a Poisson distribution assumption of photon counts.}
\end{table}

\subsection{Generalization to high-dimensional qudits}
The qudit entanglement can be generated and transferred in the same way, with the pulse sequence given by FIG.~2(d). Here, the write beam is divided equally into four paths. Prior to the detection of a signal photon, the entanglement state of the photon and MAQM1 is
\begin{equation}
|\Phi\rangle = \frac{1}{2} \sum_{i,j=1,2} |x_i,\,y_j\rangle_{s} |x_i,\,y_j\rangle_{a_1},
\end{equation}
where the relative phases between different memory cells are again set to zero. Subsequently, the spin wave in different cells can be retrieved in arbitrary order, and can be transferred to the corresponding memory cells of MAQM2. This leads to an entangled state
\begin{equation}
|\Phi'\rangle = \frac{1}{2} \sum_{i,j=1,2} |x_i,\,y_j\rangle_{s} |x'_i,\,y'_j\rangle_{a_2}.
\end{equation}

However, the verification of this qudit entanglement is more challenging: direct quantum state tomography would require a lot more measurements. Since our main purpose is to demonstrate the transfer of quantum states, here we collect the signal photon in a fixed state to simplify the measurements. Specifically, we set the signal AODs such that the signal photon modes from the four memory cells are combined with equal weight. Upon a photon detection in the generation stage, we project the MAQM1 into a W state
\begin{equation}
|\tilde{\Phi}\rangle = \frac{1}{2} \sum_{i,j=1,2} |x_i,\,y_j\rangle_{a_1},
\end{equation}
and the state of MAQM2 after the transmission is
\begin{equation}
|\tilde{\Phi}'\rangle = \frac{1}{2} \sum_{i,j=1,2}^4 |x'_i,\,y'_j\rangle_{a_2}.
\end{equation}
Following the steps of Ref.~\cite{pu2018experimental}, we measure the W state fidelity before and after the qudit state transmission in Table~1. The fidelity only decays slightly from $(94.4\pm 1.6)\%$ to $(87.7\pm 2.9)\%$, which suggests that the quantum information is well preserved during the state transfer process.
The spin wave components in each memory cell when stored in MAQM1 and in MAQM2 are shown in FIG.~3(c) and 3(d). Note that they are not the density matrices of the atomic qudits because we do not perform complete quantum state tomography.

\section{Discussion}
In this letter, we demonstrate the generation, transmission, storage and retrieval of quantum states between two MAQMs by time-bin qubit/qudit. The experimental results confirm that the quantum information is preserved during the process.

Higher dimensional qudit entanglement and state transmission can be achieved by the same method.
However, the retrieval efficiency is mainly limited by the time to transfer the time-bin qudits, which is proportional to the dimension, compared with the memory time of the atomic ensembles. On the one hand, the interval between adjacent time bins is lower-bounded by the switch time of $2\,\mu$s of the AODs, which is governed by the acoustic speed in the AO crystal and the waist of the laser beam on the AODs. To shorten the switch time, we may use materials with higher acoustic speed and suppress the laser beam waist. On the other hand, the memory times of the spin waves in both our atomic ensembles are only tens of microseconds. It can be extended to sub-millisecond by optically pumping all the atoms to a magnetic-field insensitive state \cite{zhao2009long}.

The reduction in the state fidelity is primarily caused by the nonuniformity of the optical depth in different memory cells, thus different retrieval efficiencies. The MAQM in our experiment requires a large cross section of the ensemble for the large number of memory cells. Therefore, the cigar-shaped ensemble cannot be used even if it has almost uniform EIT retrieval efficiency \cite{PhysRevLett.120.183602,vernaz2018highly}. Nevertheless, improvement is still possible if we individually trap several ensembles in the same vacuum chamber \cite{chisholm2018three}. Besides, the number of micro-ensembles can be increased by squeezing the waist of laser beams and photon modes, as well as loading larger atomic ensembles.

\section{Methods}
\subsection{Initialization of atomic ensembles}
The $^{87}$Rb atomic cloud of MAQM1 is first cooled by a 2D$^{+}$ MOT and then by a 3D MOT. The three pairs of strong cooling beams are red detuned to the D2 cyclic transition $|5S_{1/2},\,F=2\rangle\leftrightarrow|5P_{3/2},\,F=3\rangle$, and the repumping beams are on resonance to the $|5S_{1/2},\,F=1\rangle\leftrightarrow|5P_{3/2},\,F=2\rangle$ transition. The angle between the two pairs of horizontal cooling beams is set to $60^{\circ}$ to produce an ellipsoidal ensemble with about two billion atoms, such that larger cross section can be achieved to support more micro-ensembles. Then we apply a compressed MOT for $10\,$ms by increasing the detuning of the cooling beams and the intensity of the trap coil current to twice as large. After this stage, the ensemble stays at the center of the trap, and is large enough for the experiment. The atoms are further cooled by polarization gradient cooling (PGC) for $7\,$ms, reaching a final temperature of about $25\,\mu$K, and an optical depth of about 10 for the D1 transition $|5S_{1/2},\,F=2\rangle \leftrightarrow |5P_{1/2},\,F=2\rangle$. The details about the preparation of MAQM2 is described in Ref.~\cite{Jiang2019Experimental}. The memory time of MAQM1 is about $65\,\mu$s, and that of MAQM2 is about $27.8\,\mu$s.
Before the experiment, we apply $700\,$ns pulses on the targeted micro-ensembles of MAQMs to optically pump the atoms to the ground state $|g\rangle\equiv|5S_{1/2},\,F=1\rangle$.

\subsection{Multiplexing and de-multiplexing RF signals}
The RF signals for AODs (AA DTSXY-400) are generated by arbitrary waveform generators (AWG, Tektronix 5014C). The relative phases between different optical paths are intrinsically stable for multiplexing and de-multiplexing, hence can be adjusted by varying the phases of different RF frequency components on the AODs. To form the memory cell array in MAQM1, the RF frequency of its crossed AODs is swept from $95.5\,$MHz to $103\,$MHz in the Y direction, and from $97\,$MHz to $103\,$MHz in the X direction, both with a step size of $1.5\,$MHz. As for MAQM2, the RF frequency is scanned from $99\,$MHz to $105\,$MHz in the Y direction, and from $101.1\,$MHz to $105.9\,$MHz in the X direction, both with a step size of $1.2\,$MHz.

\subsection{Laser beams for state generation, storage and retrieval}
To generate the photon-atom entanglement in MAQM1, a $100\,$ns write pulse is applied, which is blue detuned by $18\,$MHz to the D1 transition $|g\rangle\leftrightarrow|e\rangle\equiv|5P_{1/2},\,F=2\rangle$. If no signal photon is detected, a $500\,$ns clean pulse resonant to the $|e\rangle\leftrightarrow|s\rangle\equiv|5S_{1/2},\,F=2\rangle$ transition is applied to pump the atoms back to $|g\rangle$, and the process is repeated. If a signal photon is detected, we further apply the sequences in FIG.~2.

The optimal power to retrieve the spin wave from a single micro-ensemble is about $64\,\mu$W for the read and the coupling beams. It ensures the high retrieval efficiency and the bandwidth matching between the idler photon and the MAQM2. When addressing multiple cells in the cell pairs and the $2\times 2$ sub-arrays, the optimal total powers are $140\,\mu$W and $300\,\mu$W respectively. During different stages of the experiment, the powers of the read and the coupling beams on each optical path are adjusted to the appropriate values by controlling the amplitudes of RF signals in the AODs. The conditional control of the write, the read and the coupling pulses is achieved by a home-made field-programmable gate array (FPGA). It also registers the detection of the signal and the idler photons and their coincidence from the SPDs. Furthermore, it produces the event trigger for AWGs to output the next pre-programmed RF signal to the AODs.

\subsection{Timing of control pulses}
The retrieval efficiency of the collective spin wave excitation is modulated by Larmor precession of the atoms, because the background magnetic field is not completely suppressed. Therefore, we set the storage time in MAQM1 and MAQM2, and the interval between time bins, to the periods of Larmor procession for the highest efficiency. The Larmor periods of the MAQMs can be controlled by the magnetic field parallel to write/read beams in MAQM1 and the coupling beam in MAQM2; the magnetic fields are adjusted by two pairs of bias coils for each MAQM. The ambient magnetic field can be roughly compensated, so the Larmor period of MAQM1 can be widely tuned from $2\,\mu$s to $16\,\mu$s. Despite the change in the magnetic field, the ensemble remains at the center of the MOT after the compression stage, so the optical circuits need no considerable modification. The Larmor period of MAQM2 can be precisely adjusted from $1.2\,\mu$s to $1.4\,\mu$s. In the experiment of qubit (qudit) transfer, the Larmor period of MAQM1 is set to $7.8\,\mu$s ($3.9\,\mu$s), and that of MAQM2 is optimized accordingly.

\subsection{Measurement of photon-atom qubit entanglement}
To measure the signal photon and the atomic qubit (after the retrieval to a photon) in arbitrary basis $a|U\rangle + b|D\rangle$, we adjust the relative amplitude and phase of the two frequency components on the AODs that correspond to the memory cells $U$ and $D$ \cite{pu2017experimental}.
This enables us to perform the quantum state tomography using the measurement basis described in Refs.~\cite{pu2017experimental,Jiang2019Experimental}.


\textbf{Acknowledgements:} This work was supported by the Ministry of Education of China, Tsinghua University, and the National key Research and Development Program of China (2016YFA0301902).Y.K.W. acknowledges support from Shuimu Tsinghua Scholar Program and International Postdoctoral Exchange Fellowship Program (Talent-Introduction Program).

\textbf{Author Contributions:} C.L., N.J., W.C., Y.F.P, S.Z. performed the experiment under supervision of L.M.D. Y.K.W preformed theoretical analysis. C.L., Y.K.W, and L.M.D wrote the manuscript.

\textbf{Competing interests:} The authors declare no competing interests.

\textbf{Author Information:} Correspondence and requests for materials should be addressed to L.M.D.
(lmduan@tsinghua.edu.cn).

\textbf{Data Availability} The data that support the findings of this study are available
from the corresponding author upon request.

\end{document}